\documentstyle[11pt]{article}

\begin{document}

\title{A note on  the black hole remnant}

\author{ Li Xiang\\
 Center for Gravity and Relativity, Department of Physics, Nanchang University,\\ Nanchang,  330031,  Jiangxi province, P. R. China\\
 }
\date{}

\maketitle

\begin{abstract}
Analyzing the tunneling probability of a Schwarzschild black hole
with a negative log-area correction to Bekenstein-Hawking entropy, I
argue that this  correction may be closely related to a black hole
remnant. The value for the minimal black hole mass is also
discussed.

\end{abstract}
%\newpage
  Enlightened by the tunneling scenario\cite{parikh}, Arzano, Medved and Vagenas\cite{arzano} recently discussed the
 tunneling probability of a Schwarzschild black hole with the entropy
\begin{eqnarray}
S&=&\frac{A}{4}+\rho\ln A,
\end{eqnarray}
where $\rho$ is a parameter, which is determined by the model
considered. It is generally believed that $\rho$ is negative, if the
log term is attributed to the effects of  quantum
gravity.\footnote{The are some aspects  responsible for the
logarithmic correction to black hole entropy, such as string theory,
loop quantum gravity, generalized uncertainty principle and modified
dispersion relation. The relevant literatures have been listed in
Refs.\cite{arzano, page}.} However, this leads to a difficulty: the
tunneling probability
\begin{eqnarray}
\Gamma\sim\left(1-\frac{\omega}{M}\right)^{2\rho}\exp\left[-8\pi\omega\left(M-\frac{\omega}{2}\right)\right],
\end{eqnarray}
becomes divergent when  the energy of the emitted particles
approaches the black hole mass($\omega\rightarrow M$). Thus an
arbitrary big black hole  could vanish in an instant, which is
certainly unacceptable. It has been argued that the problem can be
solved by introducing a canonical correction\cite{arzano}, since it
compensates for the effect of the negative $\rho$. The motivation of
this research is  to find another possibility of regularizing the
explosive probability. As argued in the following, the black hole
remnant is a possible answer. In this note, we set $\rho=-\alpha$,
and then the corrected entropy of a Schwarzschild black hole can be
rewritten as
\begin{eqnarray}\label{entropy}
S&=&\frac{A}{4}-\alpha\ln A\nonumber\\
&=&4\pi M^2-2\alpha\ln M+const,
\end{eqnarray}
where $\alpha
>0$. Correspondingly, the tunneling probability is rewritten as
\begin{eqnarray}\label{prob}
\Gamma\sim\left(1-\frac{\omega}{M}\right)^{-2\alpha}\exp\left[-8\pi\omega\left(M-\frac{\omega}{2}\right)\right],
\end{eqnarray}
which can be bounded by imposing a stricter constraint on the energy
of the emitted particles, i.e.,
\begin{eqnarray}
\omega\leq M-M_c.
\end{eqnarray}
Since $\omega\geq 0$, the above inequality means that the black hole
mass is required to be greater than a certain scale. In other words,
 the black hole has a ground state with ``zero point" energy $M_c$.
 If so, we must answer two questions:
 what is the value for $M_c$?
how is it associated with the log-area correction?

It has been argued in Ref.\cite{adler} that  generalized uncertainty
principle may prevent the black hole evaporating
completely.\footnote{Ref.\cite{yling}  has also presented a similar
argument, in terms of a modified dispersion relation.}
 In order for the entropy to be real, the
black hole mass must satisfy $M\geq \lambda$, where $\lambda$ is a
parameter which denotes the quantum gravity correction to
uncertainty principle.\footnote{See Refs.\cite{yoneya1}--\cite{ahlu}
for the sources and details of the generalized uncertainty
principle.} The correction to Bekenstein-Hawking entropy is also
related to the parameter $\lambda$, although it is not the strict
log-area type. This enlightens us that the black hole remnant should
be closely related to the correction to entropy, if both are
attributed to quantum gravity effects.  Let us first follow the line
of argument of Ref.\cite{adler}, and make an observation on the heat
capacity of a Schwarzschild black hole with the log-area correction.
Starting from (\ref{entropy}),  we obtain the inverse temperature as
follows
\begin{eqnarray}\label{temp}
\beta=T^{-1}=\frac{dS}{dM}=8\pi M-\frac{2\alpha}{M}.
\end{eqnarray}
and then the heat capacity is given by
\begin{eqnarray}\label{heat}
C=\frac{dM}{dT}=-\beta^2\left(8\pi+\frac{2\alpha}{M^2}\right)^{-1}.
\end{eqnarray}
There is a critical mass scale
\begin{eqnarray}\label{mc1}
M_c=\sqrt{\frac{\alpha}{4\pi}},
\end{eqnarray}
at which the heat capacity vanishes. According to the argument of
Adler et al\cite{adler}, the vanishing heat capacity implies a black
hole remnant.\footnote{Recalling a system composed of large number
of harmonic oscillators, the heat capacity also vanishes when the
system is in the ground state.} The minimal mass of the black hole
may be given by (\ref{mc1}). However, this argument is more or less
speculative, because the temperature is ill defined ($T=\infty$) at
the critical point. Furthermore, the physical meanings of some
thermodynamic quantities become unclear, when the black hole
evaporation enters the last stage  dominated by quantum gravity
effects. Therefore the temperature (\ref{temp}) and the heat
capacity (\ref{heat}) may be lost to their usual senses when the
black hole mass is close to the critical value. The physical
significance of the critical mass $M_c$ should be revisited by
analyzing the black hole's tunneling probability, within the
framework of quantum mechanics.

 The necessity for the black hole remnant is also derived from the analysis of the divergent behavior of the tunneling probability.
 In fact, (\ref{prob}) is based on the following
formula\cite{parikh}
\begin{eqnarray}\label{gamma1}
\Gamma&=&|M_{fi}|^2\exp(S_f-S_i)\nonumber\\
&\sim & \exp(\Delta S)
\end{eqnarray}
where the prefactor $M_{fi}$ denotes the amplitude of the process
from an initial state $|i\rangle$ to the final state $|f\rangle$.
$\exp(S_f-S_i)$ is the phase factor, which is determined by the
difference between the entropies of the initial and final
states.\footnote{In Ref.\cite{bekenstein}, $|M_{fi}|$ was supposed
to be a constant. It has also been  shown that $|M_{fi}|$ equals
unity if WKB approximation is valid. Except this, the more details
of $M_{fi}$ are absent. So the conclusion of this note is mainly
obtained by analyzing the phase factor.}
 The formula (\ref{gamma1}) has been verified by many
efforts\cite{vagenas}--\cite{wusq2}.  Obviously, $\Gamma\rightarrow
\infty$ as $S_f\rightarrow\infty$. However, the divergence of the
entropy is attributed to two hidden assumptions:(i) the black hole
mass is arbitrary; (ii) the formula (\ref{entropy}) is always valid
for an infinitesimal black hole. Concretely speaking, the formula
(\ref{entropy}) means
\begin{eqnarray}
\lim_{M\rightarrow 0}S=\infty,
\end{eqnarray}
which leads to a trouble in explaining the origin of the explosive
entropy when the black hole vanishes. Since the entropy measures the
number of degrees of freedom, it is hard to understand that
 infinite number of degrees of freedom can be confined in an object
 whose energy is almost zero.   The simplest solution of this problem is that the black hole mass is bounded by a nonzero value.
 It also means that there is a mass scale below which  the entropy formula (\ref{entropy}) is invalid.
 In the following discussion, I argue that  a reasonable value for the
critical mass may be given by (\ref{mc1}).

 When a particle with energy $\omega$ escapes from the
black hole horizon, the tunneling probability is determined by
(\ref{prob}), and it can be rewritten as
\begin{eqnarray}
\Gamma \sim\exp({\Delta S})
\end{eqnarray}
where
\begin{eqnarray}\label{ds}
 \Delta S&=&-8\pi\omega\left(M-\frac{\omega}{2}\right)-2\alpha\ln
\left(1-\frac{\omega}{M}\right)\nonumber\\
 &=&-8\pi M^2x\left(1-\frac{x}{2}\right)-2\alpha\ln(1-x),
\end{eqnarray}
where $x=\omega/M$.  On one hand, there is a  mass scale below which
the above formula fails. On the other hand, the tunneling
probability should be bounded when the black hole mass is greater
than this scale. This implies the following inequality
\begin{eqnarray}\label{inequality1}
\Delta S=-8\pi\omega\left(M-\frac{\omega}{2}\right)-2\alpha\ln
\left(1-\frac{\omega}{M}\right)\leq 0,
\end{eqnarray}
which is certainly the simplest manner to bound the tunneling
probability. However, it is also based on such a consideration: the
black hole entropy should decrease when the particles escape from
the horizon, otherwise we are in trouble of the infinite number of
degrees of freedom as $M\rightarrow 0$. Furthermore,  the inequality
(\ref{inequality1}) ensures the positivity of the black hole
temperature.\footnote{As argued in Refs.\cite{parikh}, the
$\omega^2$ term is important for the unitary scenario of black hole
evaporation. In order to preserve this term, we should consider
(\ref{inequality1}) instead of the ansatz $dS/dM\geq 0$. }
Considering the inequality
\begin{eqnarray}
 -\ln\left(1-\frac{\omega}{M}\right)\leq\frac{\omega}{M-\omega},
 \end{eqnarray}
 we have
\begin{eqnarray}\label{inequality5}
\Delta S \leq
-8\pi\omega\left(M-\frac{\omega}{2}\right)+\frac{2\alpha\omega}{M-\omega}.
\end{eqnarray}
 When the r.h.s of (\ref{inequality5})  satisfies
\begin{eqnarray}
-8\pi\omega\left(M-\frac{\omega}{2}\right)+\frac{2\alpha\omega}{M-\omega}\leq
0,
\end{eqnarray}
we obtain
\begin{eqnarray}\label{inequality6}
\omega\leq \frac{3M-\sqrt{M^2+2\alpha/\pi}}{2}.
\end{eqnarray}
 However, (\ref{inequality1}) is  weaker than
 (\ref{inequality6}), it is necessary to consider the following inequality
\begin{eqnarray}
-8\pi\omega\left(M-\frac{\omega}{2}\right)+\frac{2\alpha\omega}{M}\leq\Delta
S\leq 0,
\end{eqnarray}
 and then
\begin{eqnarray}\label{inequality4}
\omega\leq  2M-\frac{\alpha}{2\pi M}.
\end{eqnarray}
At  first sight, (\ref{inequality6}) is very different from
(\ref{inequality4}). However, it is interesting that both
inequalities  mean the same constraint: $ M\geq
M_c=\sqrt{\alpha/4\pi}$, which ensures the positivity of $\omega$.
Substituting the value for the least mass into the inequality
(\ref{inequality1}), we obtain
\begin{eqnarray}\label{inequality7}
\Delta S=-2\alpha x\left(1-\frac{x}{2}\right)-2\alpha\ln(1-x)\leq 0,
\end{eqnarray}
where $x=\omega/M=\omega\sqrt{4\pi/\alpha}$.  Considering the
following series
\begin{eqnarray}
\ln(1-x)=-\left(x+\frac{x^2}{2}+\frac{x^3}{3}+\cdots+\frac{x^n}{n}+\cdots\right),
\end{eqnarray}
the inequality (\ref{inequality7}) becomes
\begin{eqnarray}
\Delta
S=2\alpha\left(x^2+\frac{x^3}{3}+\cdots+\frac{x^n}{n}+\cdots\right)\leq
0,
\end{eqnarray}
which is valid only if $x=0$ or $\omega=0$. This is  a dramatic
result. In other words, no particle is emitted when the black hole
mass approaches the critical scale.

Following from the third law of thermodynamics, the entropy vanishes
(or approaches a constant) when an object is in the ground state. It
is also expected that the entropy approaches a minimum when the
black hole ceases evaporating. Thus the minimal probability should
be attributed to the direct transition from black hole  to its
ground state. As an evidence,   the tunneling probability without
the log-area correction is considered in advance, which is given
by\cite{parikh}
\begin{eqnarray}
\Gamma^{*}\sim \exp(\Delta S_{BH}),
\end{eqnarray}
where
\begin{eqnarray}
\Delta S_{BH}=-8\pi\omega\left(M-\frac{\omega}{2}\right).
\end{eqnarray}
Energy conservation requires the energy of the emitted particles
satisfy $\omega\leq M$.  When the upper bound of $\omega$ is
saturated, the tunneling probability approaches a minimum as follows
\begin{eqnarray}\label{gamma2}
\Gamma^{*}\sim \exp(-4\pi M^2)=\exp(-S_{BH}),
\end{eqnarray}
 which gives two hints: (i) the black hole has a ground state  denoted by $M=0$; (ii) it becomes almost impossible that a
big black hole vanishes at an instant, since the probability of the
transition  is exponentially suppressed by the entropy of the
initial state. We should not be surprised by the first hint, because
Bekenstein-Hawking entropy  is only the contribution of the
classical gravitational action of a black hole, whereas the black
hole remnant  should be attributed to the effects of quantum gravity
at the Planck scale. The black hole mass is allowed to be
arbitrarily small if we only consider the classical gravity.

Let us come back to (\ref{inequality1}), and  consider the log-area
correction again. If the critical mass (\ref{mc1}) denotes the
ground state of black hole, the minimal tunneling probability should
be determined by the maximum energy of the emitted particles
\begin{eqnarray}\label{max}
\omega=M-M_c.
\end{eqnarray}
 It is just the case, because (\ref{max}) satisfies
\begin{eqnarray}
\frac{\partial(\Delta S)}{\partial\omega}&=&-8\pi
(M-\omega)+\frac{2\alpha}{M-\omega}=0,\nonumber\\
\frac{\partial^2(\Delta S)}{\partial\omega^2}&=&16\pi>0.
\end{eqnarray}
Correspondingly, the minimal probability is given by
\begin{eqnarray}\label{gamma3}
\Gamma &\sim & \exp\left[-4\pi
(M^2-M_c^2)-2\alpha\ln(M_c/M)\right]\nonumber\\
&=&\left(\frac{e}{M_c^2}\right)^{\alpha}\exp\left[-4\pi
M^2+2\alpha\ln M\right]\nonumber\\
&=&\left(\frac{e}{M_c^2}\right)^{\alpha}\exp(-S)\nonumber\\
 &=&\exp\left[-(S-S_0)\right]
\end{eqnarray}
where
\begin{eqnarray}
S_0=4\pi M_c^2-2\alpha\ln M_c,
\end{eqnarray}
 is a constant which is directly related to the coefficient of the
 log-area correction. Except  this constant, the formula (\ref{gamma3}) has the same form as (\ref{gamma2}), and the
probability of transition to the energy level $M_c$ is only
determined by the initial black hole entropy. This similarity
implies that $M_c$ plays the role of the ``zero point" energy of
black hole, which is exalted to a nonzero value when the negative
log-area correction to Bekenstein-Hawking entropy is considered.

In summary,  this note makes some observations on the black hole
evaporation with a negative log-area correction to
Bekenstein-Hawking entropy. $M_c$, as given by (\ref{mc1}), is a
special mass scale with some interesting properties: (i) both the
``inverse temperature" and the ``heat capacity" vanish when the
black hole mass approaches it; (ii) the black hole entropy increases
with the mass only if the latter is greater than this scale; (iii)
no particle is emitted when the black hole mass approaches $M_c$, if
the tunneling probability is always bounded by (\ref{inequality1});
(iv) $M_c$ denotes  a final state with the minimum of the transition
rate. These properties are dramatic, they should have non-trivial
significances for the black hole evolution. In my opinion, they are
the natural hints of black hole remnant.

\section*{ACKNOWLEDGMENT} This research is supported by NSF of China
(Grant Nos. 10663001 and 10373003).


\begin{thebibliography}{99}
\bibitem{parikh}M. K. Parikh and F. Wilczek, Phys. Rev. Lett. 85 (2000)
5042, hep-th/9907001; M. K. Parikh, Gen. Rel. Grav. 36 (2004) 2419,
hep-th/0405160.
\bibitem{arzano} M. Arzano, A. J. M. Medved and E. C. Vagenas, JHEP
0509 (2005) 037, hep-th/0505266.
\bibitem{page}D. N. Page, New J. Phys. 7 (2005) 203, hep-th/0409024.
\bibitem{adler}R. J. Adler, P. Chen and D. I. Santiago,  Gen. Rel.
Grav. 33 (2001) 2101, gr-qc/0106080.
\bibitem{yling} Y. Ling, B. Hu and X. Li, Phys. Rev. D73 (2006)
087702,  gr-qc/0512083.
\bibitem{yoneya1}T. Yoneya, in {\it Wandering in the fields}, (World Scientific,
1987).
\bibitem{gross}D. J. Gross and P. F. Mende,   Nucl. Phys.  B303(1988) 407.
\bibitem{yoneya2} T. Yoneya, Mod. Phys. Lett. A4 (1989) 1587.
\bibitem{gross2} D.J. Gross, Philos. Trans. R. Soc. London A329 (1989)
401, and the earlier references therein.
\bibitem{venez2} G. Venezino, CERN-TH5366/89, and the earlier references
therein.
\bibitem{konishi}K. Konishi, G. Paffuti and Provero, Phys. Lett.
B234 (1990) 276.
\bibitem{guida} R. Guida K. Konishi and P. Provero, Mod. Phys. Lett.
A6 (1991)1487.
\bibitem{strominger}A. Strominger, hep-th/9110011.
\bibitem{mende} P. F. Mende, hep-th/9210001.
\bibitem{maggiore} M. Maggiore, Phys.Lett. B304 (1993) 65.
\bibitem{garay}L. J. Garay, Int. J. Mod. Phys. A10(1995) 145, and
references cited therein.
\bibitem{adler2} R. J. Adler, D. I. Santiago, Mod. Phys. Lett. A14 (1999)
1371.
\bibitem{kempf}A. Kempf, G. Mangano and R.B. Mann,   Phys.
Rev.  D52(1995) 1108.
\bibitem{ahlu}D. V. Ahluwalia, Phys. Letts. A 275 (2000) 31.
\bibitem{bekenstein}J. D. Bekenstein and V. F. Mukhanov, Phys. Lett.
B360 (1995) 7, gr-qc/9505012.
\bibitem{vagenas} E. C. Vagenas, Phys.Lett. B559 (2003) 65,
hep-th/0209185.
\bibitem{medved} A. J. M. Medved, Phys. Rev. D66 (2002) 124009,
hep-th/0207247.
\bibitem{wbliu} W. B. Liu, Phys. Lett. B634 (2006)  541.
\bibitem{zhangjy1} J. Y. Zhang and Z. Zhao, Phys. Lett. B638 (2006)
110, gr-qc/0512153.
\bibitem{zhangjy2} J. Y. Zhang and Z. Zhao, Nucl. Phys.
B725 (2005) 173.
\bibitem{wusq1} Q. Q. Jiang, S. Q. Wu and X. Cai, Phys. Rev. D73 (2006)
064003, hep-th/0512351.
\bibitem{wusq2} S. Q. Wu, Q. Q. Jiang, JHEP 0603 (2006)  079,  hep-th/0602033.

\end{thebibliography}
\end{document}